\documentclass[12pt,a4paper]{article}
\usepackage{bbm}
\usepackage{amsmath}
\usepackage{psfig}
\usepackage{epsf}
\usepackage{graphicx}
\usepackage{psfrag}
\newcommand{\ccorr}[1]{\langle \! \langle #1 \rangle \! \rangle}
\begin{document}

\begin{flushright}
BI-TP 2000/36\\
\end{flushright}

\begin{center}
\large{\bf Using a Fermionic Ensemble of Systems 
to Determine Excited States}
\end{center}

\vspace{0.3cm}

\begin{center}
Zdzis\l{}aw Burda$^{1,2}$ and Pawe\l{} Sawicki$^{3}$
\end{center}

\vspace{0.5cm}

\centerline{$^{1}$Fakult\"at f\"ur Physik, Universit\"at Bielefeld}
\centerline{P.O.Box 100131, D-33501 Bielefeld, Germany}

\vspace{0.3cm}

\centerline{$^{2}$Institute of Physics, Jagellonian University}
\centerline{Reymonta 4, 30-059 Krak\'{o}w, Poland}
\vspace{0.3cm}

\centerline{$^{3}$ The H.Niewodniczanski Institute of Nuclear Physics}
\centerline{ul. Kawiory 26a, 30-055 Krak\'{o}w, Poland}
\vspace{0.3cm}
\vspace{0.3cm}

\begin{abstract} \normalsize \noindent
We discuss a new numerical method for the determination
of excited states of a quantum system using  
a generalization of the Feynman-Kac formula.
The method relies on
introducing an ensemble of non--interacting identical systems with a
fermionic statistics imposed on the systems as a whole, and on determining
the ground state of this fermionic ensemble by taking the large time
limit of the Euclidean kernel. Due to the exclusion principle,
the ground state of an $n$-system ensemble is realized
by the set of individual systems occupying successively 
the $n$ lowest states, all of which can therefore be 
sampled in this way. 
To demonstrate how the method works, we consider a
one--dimensional oscillator and a chain of harmonically coupled particles.

\end{abstract}

\begin{flushleft}
\end{flushleft}

\section*{Introduction}
A central object in the lattice formulation 
of field theory is the Euclidean kernel, which 
is obtained from the quantum amplitude by a Wick rotation. 
For a $d$--dimensional quantum field theory, the Euclidean kernel can be
interpreted as a partition function for a $(d+1)$--dimensional classical
statistical theory. This theory can then be put on a lattice
and simulated by Monte Carlo methods.

The Euclidean kernel contains all the information about
the quantum system, including in particular 
the excited states which in field theory
correspond to particle states. Standard methods of extracting this
information from numerical lattice data, commonly used for instance in
QCD lattice spectroscopy, are based on the analysis of the large time
behaviour of the Euclidean kernel \cite{ls}. 

In this paper, we present a new method of determining excited states by
studying the large time behaviour of the Euclidean kernel for an ensemble
of identical systems with a fermionic statistics. The trick of using
a fermionic ensemble of systems 
allows us to subtract the contribution from the
ground state and focus directly on the contribution from the lowest excited
state. The method can be recursively extended to the next excited states.

The first part of the paper contains standard material, which we recall
here mainly to introduce the notation for the second part of the paper.
In this second part, we discuss the new method and present results
obtained from applying it to the cases
of a single harmonic oscillator and of a one--dimensional chain
of coupled particles.

We write most of our formulas in a quantum mechanical framework, which
corresponds to a zero--dimensional ($d=0$) quantum field theory. 
The generalization to ($d>0$)--dimensional cases is straightforward
and does not require a modification of the method or the underlying idea,
although as we will see some practical limitations occur in this
case.

\section*{Standard methods}

The Euclidean kernel is defined as
\begin{equation}
K_\tau (q_f | q_i ) = 
\langle q_f | e^{-\tau H} | q_i\rangle
=\sum_{n=0} \phi^*_n(q_f)\phi_n(q_i) \ e^{-\tau E_n} \, ,
\label{kh}
\end{equation}
where $\phi_n(q) = \langle q | n \rangle$ are eigenvectors of the
Hamiltonian $ H | n \rangle = E_n | n \rangle $ ordered
by $E_n$, $E_0 \le E_1 \le \dots$ . The parameter $\tau$ is real and
non-negative. We set $\hbar=1$. The kernel can be expressed as a sum
over paths propagating in time from an initial position $q_i$ at some
initial time $\tau_i$ to a final position $q_f$ at some later time
$\tau_f$, $\tau = \tau_f - \tau_i \ge 0$ \cite{f,fh}. Using the composition
rule
\begin{equation}
K_{\tau+\sigma}(q_f| q_i) = 
\int dq \ K_{\tau}(q_f|q) \ K_{\sigma}(q | q_i)
\label{cr1}
\end{equation}
and applying it many times to equal time intervals $\epsilon$
($N\epsilon = \tau$), one finds
\begin{equation}
K_\tau (q_f |  q_i) = 
\int \prod_{j=1}^{N-1} d q_j \ 
\prod_{k=1}^{N} K_\epsilon (q_{k} | q_{k-1}) \, ,
\label{k}
\end{equation}
where $q_f = q_N$ and $q_i = q_0$. As it stands, the formula (\ref{k})
is of not much practical use, since the same unknown function $K$ appears
on both sides of the equation. The idea is now to use an approximation
for small $\epsilon$, replacing on the right hand side $K_\epsilon$ with
the semi-classical propagator ${\cal K}_\epsilon$. As an
example, consider a quantum mechanical system with the Hamiltonian
$H = p^2/2 + V(q)$, which describes a particle in $D$ dimensions. Then
the semi-classical propagator ${\cal K}_\epsilon$ is for small $\epsilon$
given by
\begin{equation}
{\cal K}_\epsilon (q_{k} | q_{k-1}) = 
(2\pi\epsilon)^{-D/2} \
e^{- \epsilon \Delta S_{k}}
\label{ke}
\end{equation}
where
\begin{equation}
\Delta S_{k} 
= \frac{1}{2} \Big(\frac{q_k - q_{k-1}}{\epsilon}\Big)^2  + V(q_k) \, .
\label{sk}
\end{equation}
Using this, we can approximate the kernel (\ref{k}) for finite
$\tau = N\epsilon$ by
\begin{equation}
K(q_f| q_i) \approx
\int \prod_{j=1}^{N-1} d q_j \ 
\prod_{k=1}^{N} {\cal K}_\epsilon (q_{k} | q_{k-1}) =
\int D q \ e^{- S}
\label{ka}
\end{equation}
where
\begin{equation}
D q = (2\pi\epsilon)^{-ND/2}
\prod_{j=1}^{N-1} d q_j = {\cal C}
\prod_{j=1}^{N-1} d q_j \ , \qquad
S = \epsilon \sum_{k=1}^N \Delta S_k \, .
\end{equation}
For given $N$ and $\epsilon$, the factor ${\cal C}$ is just a
global normalization. In the limit $N\rightarrow \infty$ with
$\tau = N \epsilon = \rm{const}$, the formula ({\ref{ka}) approaches
the exact one (\ref{k}).

The expression (\ref{ka}) can be viewed as a partition function for a
classical field $\{q_j\}$ on a one--dimensional lattice, $j=0,\dots,N$,
weighted by the exponent of the Euclidean action~: $D q \, e^{-S}$. 
One can simulate such an ensemble of $\{ q_j \}$ using
standard Monte Carlo techniques. It is convenient to impose periodic
boundary conditions, {\em i.e.} to set $q_0=q_N=q$, and then to sum over
$q$. The partition function for the ensemble of such periodic chains is~:
\begin{equation}
Z  = \int \prod_{j=1}^{N} d q_j  
\prod_{k=1}^{N} {\cal K}_\epsilon (q_{k}  |  q_{k-1}) =
\int_{\rm{periodic \ } q}  D q \ e^{- S} \, .
\label{z}
\end{equation}
This function serves as an approximation of the quantum mechanical
expression
\begin{equation}
\int d q \ K_\tau(q| q) = 
\int d q \ e^{-\tau E_0} \Big\{ |\phi_0(q)|^2 + 
|\phi_1(q)|^2 e^{-\tau (E_1 - E_0)} + \dots \Big\}  \, .
\label{kxx}
\end{equation}
On the one hand, the integrand $d q \ K_\tau(q|q)$ corresponds to the
probability distribution of positions $q$ in the quantum system, but on
the other hand it also describes the probability distribution of particle
positions $q_k$ in the ensemble of classical chains. For each particle of
the chain the distribution is the same because the partition function
is translationally invariant. In the large $\tau$ limit, the integrand
$d q \ K_\tau(q|q) \rightarrow dq \ e^{-\tau E_0} |\phi_0(q)|^2$
is dominated by the contribution from the ground state. 
This is commonly known as the Feynman-Kac formula.
The distribution of particle positions can be measured in Monte Carlo 
simulations of the classical chains. 
This can serve as a practical method for determining the ground
state $|\phi_0(q)|^2$ of the quantum system. 

In the same way, one can
also measure the energy of the ground state. For each time slice $k$ 
define an estimator of the classical energy~: $E_k = T_k + V_k$, where 
$V_k = V(q_k)$ and $T_k$ denote the potential and kinetic parts,
respectively. The matrix elements of the squared momentum $p^2$ are given
by $(q_{k+1}-q_{k})(q_{k}-q_{k-1})/\epsilon^2$; therefore, the kinetic
energy $T_k$ depends in principle on three time slices. 
However, we can also
introduce a one--slice operator for $T_k$ by making use of the
virial theorem, which tells us that on average
$\langle T_k \rangle =  \frac{1}{2} \langle q_k V'(q_k) \rangle$. 
The operator on the right hand side, $q_k V'(q_k)$, is a one--slice
operator and so, with these definitions, is now $E_k$. 

Denote the average
over paths weighted by (\ref{z}) by $\langle \dots \rangle$. The average
$\langle E_k \rangle$ does not depend on $k$, so we can average over
slices $E = \frac{1}{N} \sum_k E_k$ to get an estimator with a smaller
statistical error. In the large $\tau$ limit,
\begin{equation}
\langle E \rangle = 
E_0 + (E_1-E_0) e^{-\tau (E_1 - E_0)} + \dots
\label{E}
\end{equation}
approaches the ground state energy $E_0$. Since the average on the left
hand side can be measured in the ensemble of chains, we thus have a method
for computing $E_0$. 

In the MC simulations one is restricted to chains of finite length $N$.
In order to increase $\tau=N\epsilon$ for finite $N$, one has to increase
$\epsilon$. This introduces systematic errors since the deviations between
$K_\epsilon$ and ${\cal K}_\epsilon$ grow with $\epsilon$. Moreover, a
small deviation from $K_\epsilon$ may accumulate and become amplified in
the convolution of $N$ such factors in the time interval $\tau=N\epsilon$
(\ref{ka}). This error can be reduced by including higher order corrections
in $\epsilon$ to ${\cal K}_\epsilon$ (\ref{ke}) by replacing the action
(\ref{sk}) with a sort of improved action \cite{ti}~:
\begin{equation}
\Delta 
S_k \rightarrow \Delta S^{impr}_k =
\Delta S_k + \frac{1}{24} \epsilon^2 (V'(q_k))^2 + \cdots \ .
\label{improved}
\end{equation}
which can be obtained by introducing higher order corrections
from the Baker-Campbell-Hausdorff formula
in the semi-classical approximation (\ref{ke}). 
One can now make an optimal
choice of $N$ and $\epsilon$ so as to balance the systematic and statistical
errors. As it stands, the method may be applied only to the ground state
since the contributions from higher states in (\ref{kxx},\ref{E}) are
exponentially suppressed.

To determine the first excited state from the path integral approach, one
has to remove the leading contribution. In the standard method this
is done by calculating connected correlation functions for two different time
slices $k_a$ and $k_b$ that are separated by a time interval
$\Delta \tau = \tau_b-\tau_a = \epsilon (k_b -k_a)$. For any one--slice
operator ${\cal O}$ one measures the correlation function 
$\langle {\cal O}(\tau_a) {\cal O}(\tau_b) \rangle$ in the ensemble of
chains, which is related to the following expression in the quantum system~: 
\begin{equation}
\langle {\cal O}(\tau_a) {\cal O}(\tau_b) \rangle =
\frac{1}{Z} \sum_{m,n} e^{-(\tau - \Delta \tau) E_m}
\langle m | {\cal O} | n \rangle e^{-\Delta \tau E_n} 
\langle n | {\cal O} | m \rangle  \, .
\end{equation}
In the limit of large $\tau$, {\em i.e.} $\tau \gg \Delta \tau$, only the
contribution from the ground state $|m\rangle = |0\rangle$ survives in the
sum. One gets in this limit~:
\begin{equation}
\langle {\cal O}(\tau_a) {\cal O}(\tau_b) \rangle = 
\sum_{n=0}^{\infty} \big|\langle 0 | {\cal O} | n \rangle\big|^2 
e^{-\Delta \tau (E_n-E_0)} \, .
\end{equation}
The first term in the sum, $|\langle 0 | {\cal O} | 0 \rangle|^2$, is
independent of $\Delta \tau$ and can be subtracted by calculating the
connected correlator~:
\begin{equation}
\ccorr{{\cal O}(\tau_a) {\cal O}(\tau_b)} = 
\langle {\cal O}(\tau_a) {\cal O}(\tau_b) \rangle -
\langle {\cal O}(\tau_a) \rangle 
\langle {\cal O}(\tau_b) \rangle \,  .
\end{equation}
For large $\Delta \tau$ this is dominated by
\begin{equation}
\ccorr{{\cal O}(\tau_a) {\cal O}(\tau_b)} = 
|\langle 0 | {\cal O} | 1 \rangle |^2 
e^{-\Delta \tau (E_1- E_0)} + \dots \, .
\label{cc}
\end{equation}
Thus, the difference $E_1 - E_0$ can be numerically determined by measuring
the exponential fall--off of the correlation function (\ref{cc}) for large
$\Delta \tau$. The freedom one has in choosing the operator ${\cal O}$
should be used in practice to maximize the coefficent
$|\langle 0 | {\cal O} | 1 \rangle |^2$ relative
to the coefficents for higher states $n>2$.

It is clear from the discussion above that this method requires a
large separation of time scales 
$0 \ll \Delta \tau \ll \tau$. This is a strong
practical restriction. Moreover, computations of connected correlation
functions are in general much more time--consuming than calculations of
averages. An additional difficulty comes from the fact that the
signal-to-noise ratio is usually very small in the region of large
$\Delta \tau$, where one has to carry out measurements of the exponential
fall--off coefficient, $E_1-E_0$. 

To summarize, this method is much more demanding in computer power than
the procedure for determining the energy of the ground state that we
described before. 

\section*{Antisymmetrization over systems}

We now propose another way of subtracting the contribution from the ground
state. We first consider an ensemble consisting of two non--interacting
identical systems, with a fermionic statistics imposed on the systems as a
whole. The Pauli principle then forbids the two individual systems to be
in the same state simultaneously. Thus, the ground state of the
ensemble corresponds to one of the sub--systems being in its 
ground state and the other one occupying the lowest excited state. 

Denote the two systems by $Q$ and $R$. They are independent and
indistinguishable. Impose the fermionic statistics on them by defining
the propagator for the twin system, $\widehat{K}^{(2)}$, as the
antisymmetrization of the individual propagators~:
\begin{equation}
\widehat{K}^{(2)}_\tau(q_f,r_f | q_i,r_i) = \frac{1}{2!}
\left\{ K_\tau(q_f | q_i) K_\tau(r_f | r_i)  
- K_\tau(r_f | q_i) K_\tau(q_f | r_i) \right\} \, .
\end{equation}
It follows from (\ref{cr1}) that this function fulfills the same kind of
composition rule~:
\begin{equation}
\widehat{K}^{(2)}_{\tau+\sigma}(q_f,r_f | q_i,r_i) = 
\int d q d r \ \widehat{K}^{(2)}_\tau(q_f,r_f | q,r) 
\widehat{K}^{(2)}_\sigma(q,r| q_i,r_i) \, .
\end{equation}
Thus, we can repeat the same approximation scheme as for the individual 
propagators (\ref{ka}). We get
\begin{equation}
\widehat{K}^{(2)}_\tau(q_f,r_f| q_i,r_i) \approx
\int \prod_{j=1}^{N-1} d q_j d r_j 
\prod_{k=1}^{N} \widehat{{\cal K}}^{(2)}_\epsilon 
(q_{k},r_k | q_{k-1},r_{k-1}) \, ,
\label{kt}
\end{equation}
where
\begin{equation}
\widehat{\cal K}^{(2)}_\epsilon (q_{k},r_k|q_{k\!-\!1},r_{k\!-\!1})= 
\frac{1}{2!} \! \left\{ 
{\cal K}_\epsilon (q_{k}  |  q_{k\!-\!1}) 
{\cal K}_\epsilon (r_{k}  |  r_{k\!-\!1})\!-\!
{\cal K}_\epsilon (r_{k}  |  q_{k\!-\!1}) 
{\cal K}_\epsilon (q_{k}  |  r_{k\!-\!1}) \right\} .
\label{antys}
\end{equation}
Now we can define the partition function for the $QR$ ensemble with
periodic boundary conditions, analogously to (\ref{z}). The partition
function is related to the underlying quantum mechanical quantities
as follows~:
\begin{equation}
\begin{array}{ll}
\widehat{Z}^{(2)} & = 
\int d q dr \ \widehat{K}_\tau^{(2)}(q,r | q, r) \\ & \\ &
= \int d q dr \ e^{-\tau \widehat{E}^{(2)}_0} \Big\{
|\widehat{\Phi}^{(2)}_0(q,r)|^2 + 
|\widehat{\Phi}^{(2)}_1(q,r)|^2 e^{-\tau (\widehat{E}^{(2)}_1 - 
\widehat{E}^{(2)}_0)} + \dots \Big\} \\ & \\ &
\rightarrow e^{-\tau \widehat{E}^{(2)}_0} \int d q dr \
|\widehat{\Phi}^{(2)}_0(q,r)|^2 
\end{array}
\label{zt}
\end{equation}
where now $\widehat{\Phi}^{(2)}_k(q,r)$ are wave functions of the twin
system and $\widehat{E}^{(2)}_k$ are the corresponding energy levels. The
wave functions are constructed as Slater determinants of the wave
functions of the individual systems. In particular, the ground state is
$\widehat{\Phi}^{(2)}_0(q,r) = \frac{1}{\sqrt{2!}} 
\ \{ \phi_0(q) \phi_1(r) - \phi_1(q) \phi_0(r) \}$, and its energy is
given by $\widehat{E}^{(2)}_0 = E_0 + E_1$. This provides us with a
practical method of determining the lowest excited state $|\phi_1(q)|^2$
and the corresponding energy $E_1$. The probability distribution of
positions in one of the two systems, $P^{(2)}(q)$, can be expressed in
terms of the wave functions of the individual system~:
\begin{equation}
P^{(2)}(q) = \int dr \ |\widehat{\Phi}^{(2)}_0(q,r)|^2 \ =
\frac{1}{2} \Big\{ |\phi_0(q)|^2  + |\phi_1(q)|^2 \Big\}
\end{equation}
Both $P^{(2)}(q)$ and $P^{(1)}(q)=|\phi_0(q)|^2$
can be measured numerically using the MC method,
and we can put them together to find
\begin{equation}
|\phi_1(q)|^2  = 2 P^{(2)}(q) - P^{(1)}(q) \ .
\end{equation}
Similarly, by subtracting the ground state energy $E_0$ of an individual
system from the ground state energy $\widehat{E}^{(2)}_0$
of the twin system we can determine the energy 
$E_1 = \widehat{E}^{(2)}_0 - E_0$ of the first excited
state of the individual system. 

This method can be extended recursively to determine higher excited states
as well. Namely, the energy $E_k$ of the $k$-th excited state can be
computed as $E_k = \widehat{E}^{(k+1)}_0 - \widehat{E}^{(k)}_0$, the
difference of the ground state energies of the ensembles composed of $k$
and $k-1$ antisymmetrized copies of the individual system, respectively.
Similarly, for the $k$-th excited state,
\begin{equation}
|\phi_k(q)|^2 = (k+1) P^{(k+1)}(q) - k P^{(k)}(q) \, .
\end{equation}
where $P^{(k)}$ is an appropriately normalized probability distribution 
\begin{equation}
P^{(k)}(q) = \int |\widehat{\Phi}^{(k)}_0(q,r,s,\dots)|^2 \ dr ds \
\dots 
\end{equation}
for the ensemble of $k$ antisymmetrized copies. In the Monte Carlo
simulations of the ensemble of $k$ copies, the probability distribution
$P^{(k)}(q)$ is measured straightforwardly as a probability distribution
of particle positions $q$ in one of the $k$ copies.

\section*{Results}
Let us first illustrate how this method works for the one--dimensional
harmonic oscillator~: $V(q) = q^2/2$. In figure \ref{fig1} we show the
probability distributions $P^{(1)}(q)=|\phi_0(q)|^2$ and $P^{(2)}(q)$
in simulations with a single system and with the ensemble of two
antisymmetrized copies, for $N=128$ and $\epsilon=0.0625$ (corresponding
to $\tau=8$). The solid line going through the data points of $P^{(1)}(q)$
is given by the ground state function of the oscillator~:
$|\phi_0(q)|^2 = \frac{1}{\sqrt{\pi}} e^{-q^2}$. The difference of the
two numerically obtained curves $2P^{(2)}(q)-P^{(1)}(q)$ is shown in
figure \ref{fig2}, where it is compared to the function
$|\phi_1(q)|^2= \frac{2}{\sqrt{\pi}} q^2 e^{-q^2}$, which describes the
first excited state squared of the oscillator. The agreement is perfect. 

\begin{figure}
\begin{center}
\psfrag{x}{{\small $q$}}
\psfrag{y}{{\small $P(q)$}}
\includegraphics[width=9cm]{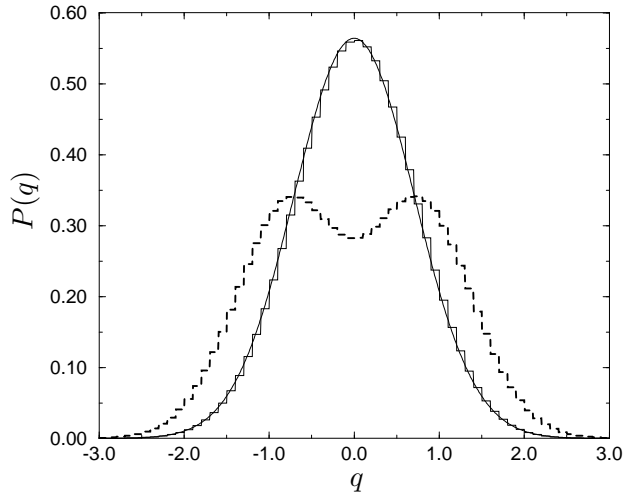}
\end{center}
\caption{\label{fig1} The histograms show the numerical results for
$P_2(q)$ and $P_1(q)$ from the MC simulations. The solid line represents
the theoretical curve $|\phi_0(q)|^2$.}
\end{figure}

\begin{figure}
\begin{center}
\psfrag{x}{{\small $q$}}
\psfrag{y}{{\small $P(q)$}}
\includegraphics[width=9cm]{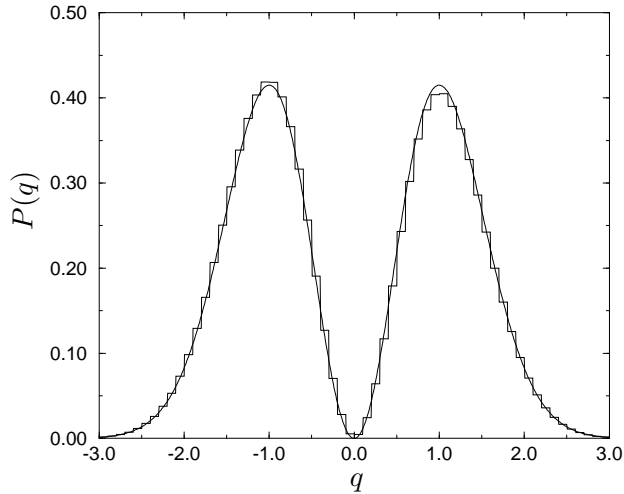}
\end{center}
\caption{\label{fig2} The histogram shows the normalized difference
$2P_2(q) - P_1(q)$ of the data from figure \ref{fig1}. The solid line
represents the theoretical curve $|\phi_1(q)|^2$.}
\end{figure}

Physically, the appearance of the valley in the $P^{(2)}(q)$ distribution
results from the fermionic repulsion between the two systems.  The
difference $|\phi_1(q)|^2=2P^{(2)}(q)-P^{(1)}(q)$ leads to a zero
value of the probability. 

The measured values of the energy of the ground
state are $E_0 = 0.501(3)$ and $E^{(2)}_0 =2.006(6)$, which gives
$E_1 = 1.505(7)$ in agreement with $E_n = n + \frac{1}{2}$. 

The method also works very well 
for higher excited states. As an example we show, in
figure \ref{fig3}, the probability distributions $P^{(4)}(q)$ and
$P^{(3)}(q)$ and, in figure \ref{fig4}, the difference
$4 P^{(4)}(q) - 3 P^{(3)}(q)$. The resulting curve is compared to the
function $|\phi_3 (q)|^2 = \frac{1}{2^3 3!} \frac{1}{\sqrt{\pi}}
H^2_3(q) e^{-q^2}$, where $H_3(q) = 8 q^3 - 12 q$ is the third Hermite
polynomial, drawn as a solid line. Again, the agreement is very good. 

As a second example, consider now a one--dimensional chain of
harmonically coupled particles with the Hamiltonian
\begin{equation}
H = \sum_j^n \left[ \frac{1}{2} p_j^2 + 
\frac{1}{2\varepsilon^2}(q_j-q_{j-1})^2 
+ V(q_j) \right]
\label{H1}
\end{equation}
The index $j$ runs over positions in the chain. The harmonic coupling
constant between neighbours is $1/\varepsilon^2$. As before, using the
path integral formulation we first write the kernel as a convolution of
$N$ propagators (\ref{k}), and then approximate
$K_\epsilon \rightarrow {\cal K}_\epsilon$ in the limit
$\epsilon\rightarrow 0$. Similarly to (\ref{ka}), we get a classical
partition function for the ensemble of fields $q_{k,j}$, distributed now
on a two-dimensional lattice and weighted by the action
\begin{equation}
S_E = \sum_{(k,j)} 
\frac{1}{2} \left(\frac{q_{k+1,j}-q_{k,j}}{\epsilon}\right)^2 + 
\frac{1}{2} \left(\frac{q_{k,j+1}-q_{k,j}}{\varepsilon}\right)^2 +
V({q}_{k,j})
\label{se}
\end{equation}
with the integration measure $\sim \prod d q_{k,j}$ (\ref{ka}),
where the index $k$ runs over time slices. Note that despite the 
apparent similarity, the two parameters $\epsilon$ 
and $\varepsilon$ have a different meaning~: 
the former is a time-slicing parameter, 
whereas the latter is a harmonic constant of the chain.
\begin{figure}
\begin{center}
\psfrag{x}{{\small $q$}}
\psfrag{y}{{\small $P(q)$}}
\includegraphics[width=9cm]{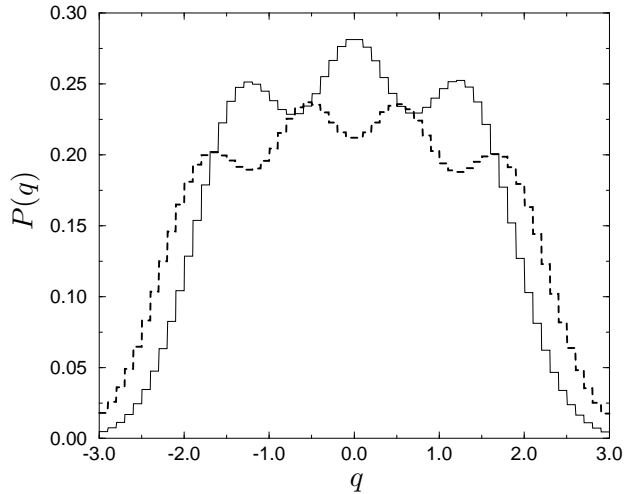}
\end{center}
\caption{\label{fig3} Numerical results for $P_4(q)$ and $P_3(q)$.}
\end{figure}
 
\begin{figure}
\begin{center}
\psfrag{x}{{\small $q$}}
\psfrag{y}{{\small $P(q)$}}
\includegraphics[width=9cm]{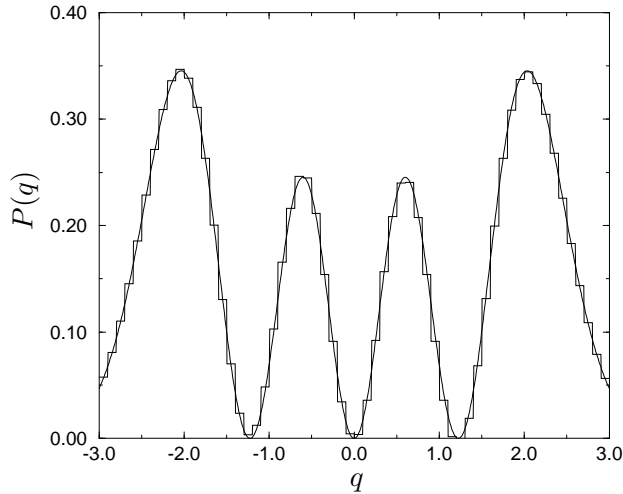}
\end{center}
\caption{\label{fig4} 
The normalized difference $4P_4(q) - 3P_3(q)$ of the data from figure
\ref{fig3}. The solid line represents the theoretical curve $|\phi_3(q)|^2$.}
\end{figure}
One way of looking at the chain is from a
quantum mechanical perspective. In this case one fixes
the harmonic constant $\varepsilon$ and for
given $\tau$ makes the time--slicing dense, $N \rightarrow \infty$
and $\epsilon=\tau/N \rightarrow 0$.
Alternatively, one can adopt a field-theoretical viewpoint
by re-interpreting the parameter $\varepsilon$ as a spatial lattice constant
and keeping it proportional to $\epsilon$~: 
$c= \varepsilon / \epsilon = const$.
In this case, the pair $(k,j)$ can be viewed as a space-time index. In
the naive continuum limit $\epsilon \rightarrow 0$, the two terms in the
sum (\ref{se}) become a Lorentz (rotationally) invariant expression
$(\partial_t q)^2 + c^2 (\partial_x q)^2$, and therefore so does the whole
theory. Thus, the expression (\ref{se}) can be regarded as a discretization
of the Euclidean action of a field theory, and the partition
function as the Euclidean kernel of a quantum field theory. We see that
the prescription for the quantization of the theory is almost the same 
as in quantum mechanics. It is commonly treated as a non--perturbative
definition of quantum field theory. 

A difference between quantum mechanics and field theory 
appears, however, when one wants to introduce the $\epsilon$ corrections. 
The formula (\ref{improved}), which works for quantum mechanics,
does not work for quantum field theory since it breaks 
the symmetry between space and time. For quantum field 
theory one therefore has to use a different 
scheme (an `improved action') to introduce lattice 
spacing corrections to the action
which preserve the Lorentz invariance of the 
underlying continuum theory \cite{ia}.

Here we will consider only the simplest case, 
namely a quantum chain with the potential $V(q) =  q^2/2$,
which we simulate on a periodic $(1+1)$-dimensional
lattice with $N$ nodes in the temporal direction 
and $n$ in the spatial one. To each node we assign 
an appropriate particle displacement $q$.  

The spectrum of this simple Hamiltonian is known. 
For convenience, we will write formulas only
for odd $n=2A+1$.
The Hamiltonian can be diagonalized in the Fourier modes $Q_\alpha$,
$\alpha=0,1,\dots,A$. The modes for $\alpha>0$ are twice degenerate,
each having a left mover and a right mover.  
The frequencies of the modes are
\begin{equation}
\omega_\alpha = 
\sqrt{1 + \frac{4}{\varepsilon^2}\sin^2 \frac{\pi \alpha}{n}} \ , 
\quad
\alpha = 1,\dots A\, .
\end{equation} 
This leads to a ground state (vacuum) energy
\begin{equation}
E_0 = \frac{1}{2} + \sum_{\alpha=1}^A \omega_\alpha
\end{equation}
and the following energies of
one--particle states, numbered by the Fourier mode index
$\alpha$,
\begin{equation}
\Delta E_{1}(\alpha) = E_{1}(\alpha) - E_0 = \omega_\alpha \, ,
\label{one_particle}
\end{equation}
where the particle mass is $m = \Delta E_{1}(0) = 1$.

For $\alpha \ll n$, the equation (\ref{one_particle}) 
reproduces the standard forumula 
$\Delta E_1 = \sqrt{m^2 + p^2}$
with the momentum $p=2\pi \alpha/(\varepsilon n)$.

The mass $m$ can be calculated as the difference between
the ground state energy of the fermionic twin system and
that of one individual system.
In practice, for finite $\tau$ such an estimator
also includes contributions
from non-zero momentum states. Although they vanish 
exponentially with $\tau$, for finite $\tau$ they
may systematically lead to results that are somewhat
overestimated.\footnote{The effect grows with $n$ 
because the energy
separation of momentum modes decreases as $n$ increases. For example,
at $\tau=5$ the maximal effect 
we measured was of order $10\%$ for $n=7$.}
To minimize this unwanted effect one should concentrate 
the analysis only on the zero momentum mode
$Q=Q_{\alpha=0}$. The effective action density for 
the zero-momentum mode is given by (\ref{sk}), 
with $q$ replaced by the zero mode $Q$.

\begin{table}
\begin{center}
\begin{tabular}{|c|c|c|c|c|c|}
\hline
         & $n = 1$  & 2        & 3        & 5       & 7       \\
\hline
$N = 16$ & 0.981(3) &          & 0.967(2) &         &         \\
32       & 1.001(3) &          & 0.993(4) &         &         \\
64       & 1.001(3) & 0.998(2) & 1.001(7) & 1.00(5) & 1.04(7) \\
128      & 0.995(3) &          & 1.02(3)  &         &         \\
\hline
\end{tabular}
\caption{The mass gap $m$ for various $N$ and $n$ ($\tau = 5$, 
$\varepsilon=1$).}
\end{center}
\end{table}
\begin{table}
\begin{center}
\begin{tabular}{|c|c|c|c|c|c|}
\hline
         & $n = 1$  & 2         & 3         & 5         & 7         \\
\hline
$N = 16$ & 1        &           & 0.2221(3) &           &           \\
32       & 1        &           & 0.0912(5) &           &           \\
64       & 1        & 0.2044(6) & 0.0436(4) & 0.0120(3) & 0.0104(3) \\
128      & 1        &           & 0.026(1)  &           &           \\
\hline
\end{tabular}
\caption{The average sign for various $N$ and $n$ 
( $\tau = 5$, $\varepsilon=1$).}
\end{center}
\end{table}

To determine the energy of $Q$, we simulate 
the whole $(1+1)$--dimensional system, estimate for finite
$\tau$ the probability that the system 
is in the zero momentum mode, and measure quantities
related to $Q$ only.
The results are summarized in tables 1-2. 

Table 1 contains the results for the particle mass. 
As we can see, the results agree with the theory; in other words,
the method works. Unfortunately, the method
is limited by the sign problem, which enters the game 
as a result of the antisymmetrization (\ref{antys}),
causing the integrand of (\ref{kt}) to not be positive 
in general. It forces us to first change the MC weights 
by taking their absolute value 
$|\widehat{{\cal K}}|$ of the integrand in (\ref{kt}),
use these modified weights in the simulations,
and then eventually to include the sign of
$\widehat{{\cal K}}$ in the estimators of the
measured quantities~:
\begin{equation}
\langle {\cal O} \rangle = 
\frac{
\langle \,
{\cal O} \ {\rm sgn} \, (\widehat{{\cal K}}) \,
\rangle_{|\widehat{{\cal K}}|}}
{\langle \,
{\rm sgn} \, (\widehat{{\cal K}}) \,
\rangle_{|\widehat{{\cal K}}|} } \, .
\end{equation}

The sole exception to this 
is the quantum system with only one degree 
of freedom, for which it is possible to show that 
the sign of the weights is always positive due to 
cancellations appearing in the two--step transfer matrix.

Generically, one expects the sign problem to create an exponential
growth in the computer demands both in $n$ and $N$, 
because the computer time required is roughly inversely 
proportional to the average sign,
which typically approaches zero exponentially.
To test this, we measured the average sign 
as a function of $n$ and $N$. As one can see
from table 2, it does indeed decrease with
growing $N$. On the other hand, as can be also seen in the table, 
for small $N$ the sign is far from zero, but then
the results for the particle mass are biased by systematic errors. 
It is a practical 
question whether there exists a window in $N$ that has
both small enough systematic errors and a large enough average sign.
This might be the case for systems which have excited states
with relatively large masses that decay over a small number
of time slices, such as for instance $SU(3)$ \cite{su3}.

\section*{Conclusions}

In conclusion, we presented a new method for the determination
of excited states in quantum mechanical systems. 
For quantum mechanics with one degree of freedom, 
the method allows to determine recursively the lowest
states and their energy. The method can be extended
to systems with more degrees of freedom, 
but in doing so one encounters the sign problem 
which in many cases can create practical limitations.
It may be that applying the recent ideas for 
defeating the sign problem \cite{sign}
to the method presented here would lead to 
an efficient method for field theoretical applications as well.

\section*{Acknowledgments}

We thank Wies\l{}aw Czy\.{z}, Bengt Petersson,
Joachim Tabaczek, and Uwe-Jens Wiese for discussions.
The work was partially supported by
the EC IHP network {\it HPRN-CT-1999-000161} and the
KBN grant 2P03B00814.

\end{document}